\documentclass[amsmath,amsfonts,aps,prl,twocolumn,showpacs]{revtex4}

\newcommand{\mati}{{j} }
\newcommand{\matmu}{{\mu} }

\newcommand{\veci}{{\bf j} }
\newcommand{\vecmu}{\mbox{\boldmath $\mu$} }
\newcommand{\up}{\uparrow}
\newcommand{\down}{\downarrow}
\newcommand{\lsf}{l_{\rm sf}}

\newcommand{\rre}{\mbox{Re}\,}
\newcommand{\iim}{\mbox{Im}\,}
\newcommand{\vq}{{\bf q}}
\newcommand{\Gratio}{g_m}
\usepackage{graphics}
\usepackage{amsmath}
\usepackage{epsf}

\begin{document}
\title{Current-induced transverse spin wave instability in a thin 
nanomagnet}
\author{M. L. Polianski and P. W. Brouwer}
\affiliation{Laboratory of Atomic and Solid State Physics,
Cornell University, Ithaca, NY 14853}

\begin{abstract}
We show that an unpolarized electric current incident
perpendicular to
the plane of a thin ferromagnet can excite a spin-wave
instability transverse to the current direction if source and
drain contacts are not symmetric. The instability,
which is driven by the current-induced ``spin-transfer torque'',
exists for one current direction only. 
\end{abstract}

\maketitle

Ferromagnets serve as spin filters for an electrical current passing
through the magnet: the spin of the electrons that are transmitted
through a ferromagnet becomes partially polarized
parallel or antiparallel to the direction
of the magnetization whereas spin current perpendicular to the
magnetization direction is absorbed. Spin filtering is the root cause
for the ``spin-transfer torque'', the phenomenon that a polarized current
impinging on a ferromagnet affects its magnetization
direction \cite{slon1,slon2,berger}. The
source of the spin polarized current can either be a different
ferromagnet, or, for a thick magnet, a region of the same ferromagnet
upstream or downstream in the current flow. The ``spin-transfer torque'' gives
rise to magnetization reversal in
ferromagnet--normal-metal--ferromagnet trilayers \cite{slon1,slon2}, 
which has been
observed experimentally by several
groups \cite{science,katine,grollier1,wegrowe,rippard,nyu,msu}. 
Dynamic manifestations of
the spin-transfer torque include domain wall motion in bulk ferromagnets 
\cite{bergerdomain,bergerexp,bazaliy,waintal} and
the excitation of spin waves by polarized currents in ferromagnetic
multilayers or wires \cite{slon2,berger,science,katine,rippard,nyu,msu,rippard2,
tsoi,ji}. 
In all these manifestations, the 
current-induced spin
torque can be distinguished from effects arising from the current
induced magnetic field, the main difference being that spin-transfer torque
effects depend on the current direction, whereas magnetic field
induced effects do not. 

In this letter, we show that an unpolarized current can also exert a 
spin-transfer torque on a ferromagnet,
even if the magnet is so thin that its
magnetization direction does not change along the current flow:
Although an unpolarized current cannot exert a spin-transfer torque that
changes
the over-all magnetization direction, it 
can create a transverse spin wave
instability for sufficiently high current densities if the source and
drain contacts to the ferromagnet are not symmetric. This spin wave
instability can be identified unambiguously as a spin-torque effect
because of its dependence on current direction: the spin-wave
instability is present for one current direction and absent for the
other. The spin-wave instability should lead to a non-hysteretic
feature in the current-voltage characteristic of the ferromagnetic
film that exists for one current direction only. In thick
ferromagnets, such features have
been observed in recent experiments \cite{science,rippard2}. 
The necessary criterion
for the spin-wave instability, asymmetric contacts to source and
drain, is generically fulfilled in experiments on nanoscale magnets
\cite{katine}.
\begin{figure}[t]
\epsfxsize=0.9\hsize
\hspace{0.05\hsize}
\epsffile{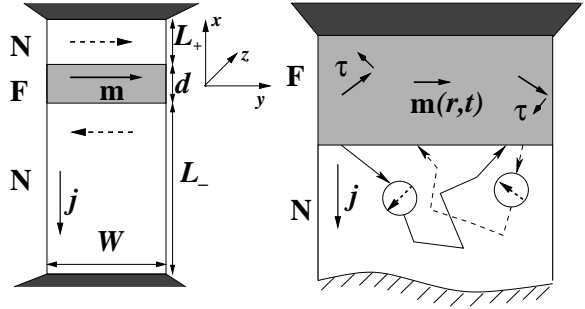}
\caption{\label{fig:1}
Left: Schematic picture of a thin ferromagnetic layer (F) in a
``nanopillar'' geometry. The ferromagnet is connected to source
and drain reservoirs through normal metal
leads (N). Current flow through the ferromagnet leads to a
spin accumulation in the normal leads, as shown by the dotted
arrows. Right: Cartoon of spin wave generation. Spin diffusion of reflected electrons 
 (diffusion paths shown full and dashed) results in a 
 spin torque on the ferromagnet's magnetization that enhances the spin wave amplitude;
  the transmitted electrons (not shown) damp the spin wave.}
\end{figure}
The issue of current-induced spin-wave excitation has significant 
practical relevance for devices based on the spin-torque effect in
ferromagnetic multilayers. Whereas, experimentally,
the presence of dynamical phenomena in these systems is well
established, the precise nature of the excitations is not well
known. A theoretical understanding of current-induced spin wave
excitation in multilayer
structures will shed light on the possibly useful application of the
current-induced dynamical excitations, {\em e.g.}, as GHz resonators, as 
well as on ways to avoid spin-wave excitation in devices that are 
designed to exhibit magnetization reversal only. For
ferromagnet--normal-metal--ferromagnet trilayers, a comparison between
theory and experiment is obstructed by the interplay of spin wave
excitations and static magnetization reversal. Our finding that
dynamical phenomena exist already in a single ferromagnetic layer
allows for a study of spin-wave excitations in a much simpler geometry
in which dynamical and static phenomena are well separated.

The geometry of the system under consideration is shown in 
Fig.\ \ref{fig:1}: a
nanomagnet of thickness $d$, small enough to be
considered as single domain, is connected to source and drain
reservoirs via diffusive
normal-metal leads of lengths $L_{-}$ and $L_{+}$,
respectively. The leads and the ferromagnet have width $W \gg d$.
In the absence of an electrical
current through the system, the ferromagnet has a uniform
magnetization, with a direction determined by anisotropies and
an external magnetic field. A current $j$ flows
perpendicular to the ferromagnet.
(Note that the electrical current $j$ points opposite
to the electron flow.)

For a qualitative
explanation of the mechanism of the spin-wave instability, we note
that the passage of an
electric current through a single ferromagnetic layer creates spin
accumulations of opposite signs on both sides of the ferromagnet, see
Fig.\ \ref{fig:1}, where the sign of the spin accumulation depends on
the current direction and on the spin-filtering properties of the
ferromagnet.  To first order in the spin-wave amplitude, 
the spin accumulation in the normal metal leads is not
affected by the possible presence of spin waves in the ferromagnet, 
as long as their wavelengths are much smaller than the
spin diffusion length in the normal metal. 
Each spin accumulation exerts a spin-transfer torque
on the magnetization; the direction of the torque is to align the
magnetization in the ferromagnet with the direction of the accumulated
spins in the leads. Depending on the sign of the spin accumulation,
such a torque either damps or enhances non-uniform
spin wave excitations in the
ferromagnet. The magnitudes of the spin accumulations on each side of
the ferromagnet are, generically, different, since they depend on the
spin diffusion length, scattering properties of
normal-metal--ferromagnet interface, distance to reservoirs, etc.
Therefore, an unpolarized current results, generically,
in a net torque that either suppresses or enhances non-uniform
spin waves in the
ferromagnet, depending on current direction. The instability occurs
when the current-induced enhancement of a spin wave amplitude
overcomes the intrinsic spin-wave damping.

For a quantitative description of the spin-wave instability, we choose
the $x$ axis along the leads, with the origin such that the
normal-metal--ferromagnet interfaces are at $x = \pm d/2$, see
Fig. \ref{fig:1}.  We assume that the ferromagnet is so thin that its
magnetization direction varies with respect to the transverse
coordinates $y$ and $z$ only, but not with respect to $x$. In the
diffusive normal-metal leads, the electron distribution is
described by potentials $\matmu_e({\bf r},t)$ for the electron density
and ${\vecmu}_s({\bf r},t)$ for the electron 
spin \cite{valetfert,brataas,huertas}. The current is
separated into the charge current density $\mati_{\alpha}$,
$\alpha=x,y,z$, and the spin current density ${\bf j}_{s\alpha}$. Since electrons adjust to the changing magnetization
on a time scale much faster than that of the magnetization dynamics,
current and potential are related by means of the time-independent
diffusion equation in the normal metal leads,
\begin{eqnarray}\label{eq:iv}
  && \nabla^2\mu_{e} = 0,\ \ 
  \mati_{\alpha} = (\sigma/e) \nabla_\alpha\matmu_{e},\ \
  \alpha=x,y,z,
  \nonumber \\
  && \lsf^2\nabla^2{\vecmu}_s = \vecmu_s,\ \
  {\bf\mati}_{s\alpha} = - 
  (\hbar \sigma/2 e^2) \nabla_{\alpha}\vecmu_{s},
\end{eqnarray}
supplemented by boundary conditions for the source and drain
reservoirs $\matmu_e(-L_{-})=-eV$,
$\matmu_e(L_{+})=0$, ${\vecmu}_s(\pm L_{\pm}) ={\bf 0}$. Here $\sigma$ is the
conductivity of the normal metal leads and $\lsf$ the spin-diffusion
length. At the normal-metal--ferromagnet interfaces, the charge and
spin current $j_x$ and ${\bf j}_{sx}$ in the normal metal perpendicular to
the interface are related to the potential drop $\Delta \mu$ over the
interface as \cite{brataas,tbb}
\begin{eqnarray}
\label{eq:discont}\label{eq:current}
 ({\bf j}_{sx})_{\perp} &=&(\hbar/2e^2)\rre g_{\up\down}
   \left(2 \Delta\vecmu_s \times {\bf m}\pm\hbar \partial_t {\bf
   m}\right)\times{\bf m}
   \nonumber \\ && \mbox{} + (\hbar/2e^2)\iim g_{\up\down}\left(2\Delta\vecmu_s 
   \times {\bf m}\pm
   \hbar \partial_t {\bf
   m}\right)
  , \nonumber\\
  ({\bf j}_{sx})_{\parallel} &=&
  -(g_{\up\up}+g_{\down\down}) 
  (\hbar/2 e^2) {\bf m} \cdot \Delta {\vecmu_s}
   \\ && \mbox{}
  - (g_{\up\up}-g_{\down\down}) 
  (\hbar/2 e^2) \Delta \matmu_e, 
  \nonumber \\
  j_{x} &=&
  (g_{\up\up}+g_{\down\down}) (\Delta \matmu_e/e)
  + (g_{\up\up}-g_{\down\down}) 
  {\bf m} \cdot (\Delta {\vecmu}_s/e). \nonumber
\end{eqnarray}
Here ${\bf m}({\bf r},t)$ is the unit vector pointing in the direction of the
magnetization of the ferromagnet, $({\bf j}_{sx})_{\perp}$ and $({\bf
j}_{sx})_{\parallel}$ are the $x$-components of the spin current
perpendicular and parallel to ${\bf m}$, respectively, $g_{\up\up}$ and
$g_{\down\down}$ are interface conductivities for spins aligned
parallel and antiparallel to ${\bf m}$, whereas $g_{\up\down}$ is
the ``mixing conductivity'' \cite{brataas,tbb}, 
$\Delta \mu = \mu(\pm d/2 + 0) -
\mu(\pm d/2 - 0)$ is the potential drop over the interface. (We assume
identical conductivities and spin-flip lengths in the leads, and
identical scattering properties of the ferromagnet--normal-metal
interfaces, although our results are readily generalized to
unequal values of $\sigma$, $l_{\rm sf}$ and $g_{\alpha\beta}$.)
At the normal-metal--ferromagnet
interface, $(\veci_{sx})_{\parallel}$ and $\mati_x$ are continuous,
whereas $(\veci_{sx})_{\perp} = 0$ in the ferromagnet. We assume that the ferromagnet is
so thin that all potential drops occur at the interfaces, so that we
can neglect the $x$-dependence of the potentials in the
ferromagnet. Then Eqs.\ (\ref{eq:iv}) and (\ref{eq:current}) fully determine
the potentials and currents in the normal metal and the ferromagnet,
as a function of ${\bf m}$.

For a thin ferromagnet, variations of the magnetization direction
${\bf m}$ along the
$x$ axis will have a large energy cost and, hence, a large
threshold for excitation. Therefore, we consider transverse
modulation of ${\bf m}$ only and take ${\bf m}$
independent of $x$. The dynamics of ${\bf m}(y,z,t)$ is 
determined by the
Landau-Lifschitz-Gilbert equation (without an applied magnetic
field \cite{field}) \cite{slon2,bazaliy},
\begin{eqnarray}\label{eq:LLG}
  \partial_t{\bf m} &=& \alpha {\bf m}\times\partial_t{\bf m}
  + J \gamma M\nabla^2 {\bf m} \times{\bf m}
  \nonumber \\ && \mbox{}
  - (\gamma/M) (K_1 m_1\hat e_1+K_2 m_2\hat e_2)\times{\bf m}
   \\ && \mbox{}
  + (\gamma/Md)[\veci_{sx}(-d/2-0) - \veci_{sx}(d/2+0)],
  \nonumber
\end{eqnarray}
where $\alpha$ and $J$ are the bulk Gilbert damping coefficient and
spin stiffness (exchange constant) respectively, 
$\gamma = \mu_B g/\hbar$ is the
gyromagnetic ratio, $M$ is the magnetization per unit volume,
and $K_1$ and $K_2$ are anisotropy
constants along principal directions $\hat e_1$ and $\hat e_2$,
respectively, obtained by expanding the magnet's free energy around 
a preferred axis $\hat e_3$. The magnetization satisfies the
boundary condition $(\hat n \cdot\vec \nabla ){\bf m} = 0$, where
$\hat n$ is the normal to the ferromagnet's surface. 
We neglect the effect of the current-induced magnetic field on the
magnetization dynamics, which is allowed if the width $W$ is
sufficiently small, $\ll 1$ $\mu$m for typical experimental parameters.
The parameters in the Landau-Lifschitz-Gilbert equation define a
length scale $1/q_{\rm f}$ and current scale $j_{\rm f}$,
\begin{equation}
  q_{\rm f}^{2} = \frac{K_1+K_2}{2 J M^2}, \ \
  j_{\rm f}^2 = \left( \frac{2e}{\hbar} \right)^2 
  J M^2\frac{K_1 + K_2}{2}.
  \label{eq:qf}
\end{equation}
The quantities $1/q_{\rm f}$ and $\hbar j_{\rm f}/e$ are proportional to 
the width and energy of a domain wall, respectively.
Order-of-magnitude estimates of the various parameters involved
here are $d \sim 10$ nm, $W \sim 10^{2}$ nm,
$\lsf$(Cu)$\sim 10^2$ nm \cite{katine,albert}; $\iim g_{\up\down}
\ll \rre g_{\up\down} \sim g_{\up\up} \sim
g_{\down\down} \sim 10^{14}$ $\Omega^{-1}$m$^{-2}$ 
\cite{stiles,xia} for Co/Cu and Fe/Cr interfaces; 
$\sigma/\lsf$(Cu)$\sim 10^{15}$ $\Omega^{-1}$m$^{-2}$, and typically
$q_{\rm f} \sim 10^{-1}$ nm$^{-1}$, 
$j_{\rm f} \sim 10^{8}$ A$/$cm$^2$  for Co, Fe or Ni \cite{reference}. 

We first solve these equations for a uniform
magnetization, ${\bf m}$ independent of the transverse coordinates
$y$ and $z$. The spin accumulation ${\vecmu}_s$ in the normal metal
leads close to the normal-metal--ferromagnet interface is
\begin{eqnarray}\label{eq:stable}
  \vecmu_s(\pm d/2) &=& \mp (e\mati_{x}/\Gratio)
  \tanh(L_{\pm}/\lsf) {\bf m},
\end{eqnarray}
where $j_x$ is the charge current density and
\begin{eqnarray*}
  \Gratio &=&
  \frac
  { (\sigma/\lsf)(g_{\up\up} + g_{\down\down}) + 2 g_{\up\up} g_{\down\down}
  \sum_{\pm} \tanh(L_{\pm}/\lsf) }
  {g_{\up\up} - g_{\down\down}}.
  \label{eq:C}
\end{eqnarray*}
The spin accumulation is shown schematically in Fig.\ \ref{fig:1}.

In the case of uniform magnetization, the spin accumulation ${\vecmu}_s$
is always
parallel to ${\bf m}$, and no current-induced torque is applied to 
the magnetization. The situation changes if ${\bf m}$ varies in the
transverse direction. In this case, transverse diffusion of spin
in the normal-metal leads gives rise to an angle between ${\vecmu}_s$
and ${\bf m}$, and, hence, to a current-induced torque. In order to
study this scenario in detail, we analyze Eqs.\
(\ref{eq:iv})--(\ref{eq:LLG}) for a small deviation of ${\bf m}$
from the equilibrium direction $\hat e_3$. The result
can be represented in terms of an equation of motion for
$\delta{\bf m} = {\bf m} - \hat e_3$. We assume a rectangular
cross section of dimensions $W_y$ and $W_z$ in the $y$ and $z$
directions and perform a Fourier transform with respect to the
transverse coordinates $y$ and $z$. The allowed wavevectors are $q_y =
\pi n_y/W_y$, $q_z = \pi n_z/W_z$, where $n_y$ and $n_z$ are
non-negative integers. Representing $\delta {\bf m}$ through  
$m_{\pm}(\vq) = \delta m_1(\vq) \pm i \delta m_2(\vq)$, one finds 
that, to first order in $m_{\pm}$, the equations
of motion of different Fourier modes separate,
\begin{eqnarray}
  \label{eq:LLGnew}
  \left(\frac{\tilde\alpha}{\gamma} \pm \frac{i}
  {\tilde \gamma} \right) M \partial_t m_{\pm}
  &=& \frac{K_2-K_1}{2} m_{\mp}
  - \frac{\hbar j_{\rm f}(q^2 + q_{\rm f}^2)}{2 e q_{\rm f}} m_{\pm} 
  \nonumber
   \\ &&   \mbox{} -
  \frac{\hbar \mati_{x}(S_2\mp iS_1 )}{2 e d}
   m_\pm,
\end{eqnarray}
Here $\tilde \alpha$ and $\tilde
\gamma$ are renormalized Gilbert damping
parameter and gyromagnetic ratio,
\begin{subequations} \label{eq:ag}
\begin{eqnarray}
\frac{1}{\tilde \gamma} &=&
  \frac{1}{\gamma} + 
  \frac{\hbar^2}{2Mde^2}\iim \sum_{\pm}\frac{g_{\up\down}G_{\pm}(q)}
  {G_{\pm}(q)+g_{\up\down}} ,
  \label{eq:gamma}\\
  \label{eq:alpha}
  \tilde\alpha&=&\alpha+\frac{\gamma\hbar^2}{2Mde^2}
  \rre\sum_{\pm}\frac{g_{\up\down}G_{\pm}(q)}
  {G_{\pm}(q)+g_{\up\down}},
\end{eqnarray}
\end{subequations}
whereas the dimensionless numbers $S_1$ and $S_2$ set the
magnitude of the current-induced torque,
\begin{subequations}
\begin{eqnarray}
  \label{eq:LLGnewS1}
  S_1 &=& \frac{\sigma}{\Gratio \lsf}\rre\sum_{\pm}
  \frac{ \pm g_{\up\down}}{G_{\pm}(0)}
  \frac{G_{\pm}(0)-G_{\pm}(q)}{g_{\up\down}+G_{\pm}(q)},\\
  S_2 &=&\frac{\sigma}{\Gratio\lsf }\iim\sum_{\pm}
  \frac{ \pm g_{\up\down}}{G_{\pm}(0)}
  \frac{G_{\pm}(0)-G_{\pm}(q)}{g_{\up\down}+G_{\pm}(q)},
\end{eqnarray}
\end{subequations}
where $\Gratio$ was defined below Eq.\ (\ref{eq:stable}) and
\begin{eqnarray*}
  G_{\pm}(q)&=& \frac \sigma 2\sqrt{\lsf^{-2}+q^2}
  \coth \left(L_{\pm} \sqrt{\lsf^{-2}+q^2}\right).
\end{eqnarray*}
In the limit $q \to 0$, Eqs.\ (\ref{eq:ag}) coincide with
the enhanced Gilbert damping and gyromagnetic ratio reported by
Tserkovnyak {\em et al.} \cite{tbb}.

In the absence of a current, any spatial modulation of the
magnetization is damped.  It is the existence of the source terms
$S_1$ and $S_2$ in Eq.\ (\ref{eq:LLGnew}) that leads to a spin-wave
instability at sufficiently large current density $\mati$.  Note
that the source terms exist only if the normal leads are asymmetric
(different lengths, or different conductivities, spin-flip lengths or
interface conductivities), and if the ferromagnet is a spin filter,
$g_{\up\up} \neq g_{\down\down}$. The source term $S_2$ gives rise to
a small change of the ferromagnetic resonance frequency, whereas $S_1$
increases or decreases the amplitude of the spin wave, depending on
the current direction. An instability occurs if the current-induced
enhancement of the spin-wave amplitude overcomes the damping, i.e., if
\begin{eqnarray}\label{eq:instab}
  \left( \frac{\gamma S_1}{\tilde \alpha \tilde \gamma} - 
  S_2\right)(-\mati) > \frac{(q^2 + q_{\rm f}^2)d}{q_{\rm f}}j_{\rm f}.
\end{eqnarray}

\begin{figure}[t] 
\epsfxsize=0.80\hsize
\hspace{0.05\hsize}
\epsffile{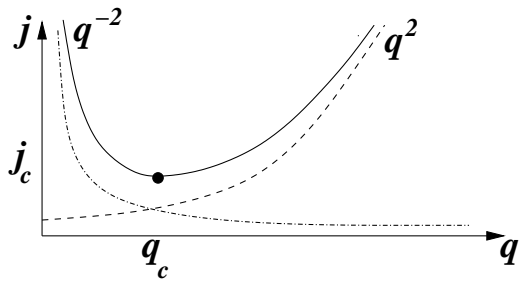}
\caption{\label{fig:2}
Schematic picture of the current density $\mati_{e}$ required to
excite a spin-wave at wavevector $\vq$. The
spin-wave instability occurs at the wavevector $\vq$ for which
$\mati$ is minimal.}
\vspace{-0.2cm}
\end{figure}

Using the order-of-magnitude estimates listed below Eq.\
(\ref{eq:qf}), the instability criterion simplifies considerably
if we set $\iim g_{\up\down} = 0$, take the limits $d \to 0$ and
$\sigma/\lsf \gg \rre g_{\up\down}$, and
consider the case $L_{-} \gg \lsf \gg L_{+}$ of maximally asymmetric
contacts,
\begin{eqnarray}
  \frac{-\mati}{j_{\rm f}} &>&
  \frac{\hbar^2\gamma g_m}{Mq_{\rm f}e^2}\frac{q_{\rm f}^2+q^2}
  {1-(1+q^2\lsf^2)^{-1/2}}.
  \label{eq:icrit}
\end{eqnarray}
The r.h.s.\ of Eq.\ (\ref{eq:icrit}) is shown schematically in Fig.\
\ref{fig:2}. For large $q \gg 1/\lsf$, the spin-wave instability is
dominated by the stiffness of the spin wave, which leads to a critical
current density $\propto q^2$. For small $q \ll 1/\lsf$, the magnetization
accumulation in the normal metal leads tends to remain locally
parallel to the magnetization, so that the current-induced torque is
strongly reduced and the
critical current density for excitations at small $q$
is increased $\propto q^{-2}$.
For a thick ferromagnet,
the effect of bulk Gilbert damping becomes dominant, causing the
critical current density to increase linearly
with the thickness $d$.

 In the 
experimentally relevant parameter regime 
$1/q_{\rm f} \ll \lsf \ll \sigma/\rre g_{\up\down}$, 
the critical current density is 
\begin{eqnarray} \label{eq:wide}
  \mati_{\rm c} &=& \frac{\hbar^2\gamma \Gratio q_{\rm f}}{Me^2}j_{\rm f},
  \ \  q_{\rm c} = \left(q_{\rm f}^2/2\lsf\right)^{1/3}.
\end{eqnarray}
For a Cu/Co/Cu geometry, we estimate $\tilde\alpha \sim (0.1$nm$)/d$
and $\hbar^2 \gamma \Gratio q_{\rm f}/M e^2 \sim 0.2$, as long as 
$d q_{\rm f}
\lesssim 1$. We then find $\mati_{\rm c}\sim 10^7$ A$/\mbox{cm}^2$,
which is of the same order of magnitude as the critical current
for magnetization reversal in a ferromagnet--normal-metal--ferromagnet
trilayer\cite{katine}. The estimate (\ref{eq:wide}) is valid as 
long as the width
$W \gg 1/q_{\rm c}$. For small $W$, the instability occurs
at the lowest possible wavenumber, $q_{\rm c} = \pi/W$.


In conclusion, we have shown that an unpolarized electric current
passing through a thin ferromagnetic layer can create a 
spin wave instability with wavevector transverse
to the current. The instability occurs at one current direction
only and depends on the asymmetry of normal metal contacts. The
mechanism for the instability is the same as the one believed to
cause magnetization reversal in ferromagnetic multilayers,
{\em i.e.,} a
spin-transfer torque arising from 
spin accumulation in the normal metal contacts
perpendicular to the magnetization direction ${\bf m}$ 
and spin-filtering at the
normal-metal--ferromagnet interfaces \cite{slon1,slon2,bazaliy}.

Current-induced excitations that occur for one current direction 
only have been observed in single ferromagnetic
layers \cite{science,ji,rippard2}. For ferromagnets much
thicker than the inverse wavevector $q_{c}^{-1} \sim 3 \times
10^{1}$nm for transverse spin waves, cf.\ Eq.\
(\ref{eq:wide}), spin wave excitations with wavevector along the
current flow may have a lower threshold current than transverse
excitations \cite{bazaliy,ji}. Such thick ferromagnetic layers
are common in the point-contact geometry 
\cite{science,rippard,ji}.
It is for the thin layers used in the
nanopillar geometry \cite{katine,nyu,msu,grollier1,albert} 
that we believe the transverse instability
considered here is most significant.

We thank Yaroslav Bazaliy, Nathan Emley, Sergey Kiselev, Ilya
Krivorotov, Dan Ralph, Jack Sankey and Yaroslav Tserkovnyak 
for discussions. This work was supported
by the Cornell Center for Materials research under NSF grant no.\
DMR 0079992, the Cornell Center for Nanoscale Systems under NSF 
grant no.\ EEC-0117770, the NSF under grant no.\ DMR 0086509, and the
Packard Foundation.

\newpage


\begin{thebibliography}{66}\vspace{-0.3cm}

\bibitem{slon1}
J. Slonczewski, J. Magn. Magn. Mater. {\bf 159}, L1 (1996).

\bibitem{slon2}
J. Slonczewski, J. Magn. Magn. Mater. {\bf 195}, L261 (1999)

\bibitem{berger}
L. Berger, Phys. Rev. B {\bf 54}, 9353 (1996).

\bibitem{science}
E.~B.~Myers {\em et al.},
Science {\bf 285}, 867 (1999).

\bibitem{katine}
J. A. Katine {\em et al.}, 
Phys.~Rev.~Lett. {\bf 84}, 3149 (2000).


\bibitem{rippard}
W.~H.~Rippard, M.~R.~Pufall, and T.~J.~Silva,
Appl.~Phys.~Lett. {\bf 82}, 1260 (2003).

\bibitem{nyu}
J.\ Z.\ Sun {\em et al.}, Appl.\ Phys.\ Lett.\ {\bf 81}, 2202
(2002).


\bibitem{msu}
S.~Urazhdin {\em et al.}, Phys.\ Rev.\ Lett.\ {\bf 91}, 146803 (2003).
\bibitem{grollier1} J.\ Grollier {\em et al.},
Appl.\ Phys.\ Lett.\ {\bf 78}, 3663 (2001).

\bibitem{wegrowe}
J.-E. Wegrowe {\em et al.}, Appl. Phys. Lett. {\bf 80}, 3775 (2002).

\bibitem{bergerdomain}
L. Berger, Phys. Lett. A {\bf 46A}, 3 (1973).

\bibitem{bergerexp} P. P. Freitas and L. Berger, J.\ Appl.\ 
Phys.\ {\bf 57}, 1266 (1985); C.-Y. Hung and L. Berger, J. Appl.\
Phys.\ {\bf 63}, 4276 (1988).

\bibitem{bazaliy} Ya.~B.~Bazaliy, B.~A.~Jones, and S.-~C.~Zhang, Phys.~Rev.~B
{\bf 57}, R3213 (1998).

\bibitem{waintal} X. Waintal and M. Viret, cond-mat/0301293.


\bibitem{tsoi}
M.~Tsoi {\it et al.}, Nature {\bf 406}, 46 (2000).

\bibitem{ji}
Y. Ji, C. L. Chien, and M. D. Stiles, Phys. Rev. Lett.
{\bf 90}, 106601 (2003).

\bibitem{rippard2} W. H. Rippard, M. R. Pufall, and T. J. Silva,
unpublished.



\bibitem{valetfert} T. Valet and A. Fert, Phys.\ Rev.\
B {\bf 48}, 7099 (1993).

\bibitem{brataas} A.~Brataas, Yu.~V.~Nazarov, and G.~E.~W.~Bauer, 
Phys.~Rev.~Lett.\ {\bf 84}, 2481 (2000).

\bibitem{huertas} D.\ Huertas-Hernando {\em et al.}, Phys. Rev. B
{\bf 62}, 5700 (2000).

\bibitem{tbb} Ya.~Tserkovnyak, A.~Brataas, and G.~E.~W.~Bauer, 
Phys.~Rev.~Lett. {\bf 88}, 117601 (2002);
Phys.~Rev.~B {\bf 66}, 224403 (2002).

\bibitem{field}
The effect of a magnetic field is mathematically equivalent to a
change of the equilibrium magnetization direction $\hat e_3$
and the anisotropy constants
$K_1$ and $K_2$. A magnetic field parallel (antiparallel) to the
easy axis increases (decreases) $K_1$ and $K_2$ and, hence, shifts the 
spin-wave instability to higher (lower) current densities. 

\bibitem{albert}
F.~J.~Albert {\em et al.},
Phys.~Rev.~Lett. {\bf 89}, 226802 (2002).
\bibitem{stiles}
M. D. Stiles, J. Appl. Phys. {\bf 79}, 5805 (1996); Phys.
Rev. B. {\bf 54}, 14679 (1996).

\bibitem{xia}
K. Xia {\em et al.},
Phys.~Rev.~B {\bf 65}, 220401 (2002).


\bibitem{reference} 
See E.\ P.\ Wohlfahrt in E.\ P.\ Wohlfahrt, Ed., {\em Ferromagnetic Materials}, Vol.\
1 (North-Holland, 1980) for material constants. For a thin magnet the shape 
anisotropy dominates over the bulk easy axis anisotropy, $K_1 \approx 4\pi M\gg K_2$.

\end{thebibliography}
\end{document}